\documentclass[twocolumn,showpacs,aps,prl,letterpaper]{revtex4}
\usepackage{graphicx}
\usepackage{dcolumn}
\usepackage{amsmath}
\usepackage{epsfig}
\usepackage{eepic}
\usepackage{color}

\allowdisplaybreaks

\long\def\inst#1{\par\nobreak\kern 4pt\nobreak
    {\itshape #1}\par\vskip 10pt plus 3pt minus 3pt}
\RequirePackage{xspace}

\def\babar{\mbox{\slshape B\kern-0.1em{\smaller A}\kern-0.1em
    B\kern-0.1em{\smaller A\kern-0.2em R}}}
\def\Kbar    {\kern 0.18em\overline{\kern -0.18em K}{}\xspace}

\def\Kz      {\ensuremath{K^0}\xspace}
\def\Kzb     {\ensuremath{\Kbar^0}\xspace}
\def\KzKzb   {\ensuremath{\Kz {\kern -0.16em \Kzb}}\xspace}

\def\Ks     {\ensuremath{K_S}\xspace}
\def\Kl     {\ensuremath{K_L}\xspace}
\def\KsKs   {\ensuremath{\Ks {\kern -0.16em \Ks}}\xspace}
\def\KlKl   {\ensuremath{\Kl {\kern -0.16em \Kl}}\xspace}
\def\KsKl   {\ensuremath{\Ks {\kern -0.16em \Kl}}\xspace}
\def\KlKs   {\ensuremath{\Kl {\kern -0.16em \Ks}}\xspace}
\def\Dbar    {\kern 0.18em\overline{\kern -0.18em D}{}\xspace}
\def\Dz      {\ensuremath{D^0}\xspace}
\def\Dzb     {\ensuremath{\Dbar^0}\xspace}
\def\DzDzb   {\ensuremath{\Dz {\kern -0.16em \Dzb}}\xspace}
\def\Lam      {\ensuremath{\Lambda}\xspace}
\def\Lamb      {\ensuremath{\overline{\Lambda}}\xspace}
\def\LLb   {\ensuremath{\Lam {\kern -0.16em \Lamb}}\xspace}

\newcommand{\DsP}{\ensuremath{D_s^+}\xspace}
\newcommand{\DsM}{\ensuremath{D_s^-}\xspace}
\newcommand{\DspDsm}{\ensuremath{\DsP {\kern -0.16em \DsM}}\xspace}
\newcommand{\Dp}{\ensuremath{D^+}\xspace}
\newcommand{\Dm}{\ensuremath{D^-}\xspace}
\newcommand{\DpDm}{\ensuremath{\Dp {\kern -0.16em \Dm}}\xspace}

\def\Bbar    {\kern 0.18em\overline{\kern -0.18em B}{}\xspace}

\def\Bz      {\ensuremath{B^0}\xspace}
\def\Bzb     {\ensuremath{\Bbar^0}\xspace}
\def\BzBzb   {\ensuremath{\Bz {\kern -0.16em \Bzb}}\xspace}
\def\Bu      {\ensuremath{B^+}\xspace}
\def\Bub     {\ensuremath{B^-}\xspace}

\def\BpBm    {\ensuremath{\Bu {\kern -0.16em \Bub}}\xspace}

\def\Dp      {\ensuremath{D^+}\xspace}

\newcommand{\optbar}[1]{\shortstack{{\tiny (\rule[.4ex]{1em}{.1mm})}
  \\ [-.7ex] $#1$}}
\def\BorBbar    {\kern 0.18em\optbar{\kern -0.18em B}{}\xspace}
\def\DorDbar    {\kern 0.18em\optbar{\kern -0.18em D}{}\xspace}
\def\KorKbar    {\kern 0.18em\optbar{\kern -0.18em K}{}\xspace}

\def\pep2{PEP-II}
\mathchardef\Upsilon="7107
\def\Y#1S{\ensuremath{\Upsilon{(#1S)}}\xspace}% no space before {...}!

\begin{document}

\title{\large \bfseries \boldmath  Probe $\Lam - \Lamb$ oscillation in $J/\psi \rightarrow \Lambda\,\overline\Lambda$ decay
        at BES-III }
\author{Xian-Wei Kang$^{1,2}$}\email{kangxw@ihep.ac.cn}
\author{Hai-Bo Li$^1$}\email{lihb@ihep.ac.cn}
\author{Gong-Ru Lu$^2$}\email{lugongru@sina.com}
\affiliation{$^1$Institute of High Energy Physics, P.O.Box 918,
Beijing  100049, China\\ $^2$Department of Physics, Henan Normal
University, Xinxiang 453007, China}

%%%%%%%%%%%%%%%%%%%%%%%%%%%%%%%%%%%%%%%%%%%%%%%%%%%%%%%%%%%%%%%%%%%%%%%%%%%%%%%%%%
%                             DATE                                              %%
%%%%%%%%%%%%%%%%%%%%%%%%%%%%%%%%%%%%%%%%%%%%%%%%%%%%%%%%%%%%%%%%%%%%%%%%%%%%%%%%%%

%\date{\today}
%\date{March 32, 2003}

%%%%%%%%%%%%%%%%%%%%%%%%%%%%%%%%%%%%%%%%%%%%%%%%%%%%%%%%%%%%%%%%%%%%%%%%%%%%%%%%%%
%                             Abstract                                          %%
%%%%%%%%%%%%%%%%%%%%%%%%%%%%%%%%%%%%%%%%%%%%%%%%%%%%%%%%%%%%%%%%%%%%%%%%%%%%%%%%%%

\begin{abstract}
We discuss the possible searching for the $\Lam - \Lamb$ oscillation
by coherent $\Lam\Lamb$ production in J/$\psi \rightarrow \Lam
\Lamb$ decay process. The sensitivity of measurement of $\Lam
-\Lamb$ oscillation in the external field at BES-III experiment is
considered. These considerations indicate an alternative way to
probe the $\Delta B =2$ amplitude in addition to neutron oscillation
experiments. Both coherent and time-dependent information can be
used to extract $\Lam-\Lamb$ oscillation parameter. With one year's
luminosity at BES-III, we can set an upper limit of $\delta m <
10^{-15}$ MeV at 90\% confidence level, corresponding to about
$10^{-6}$ s of $\Lam-\Lamb$ oscillation time.

\end{abstract}

\pacs{11.30.Fs, 12.60.Cn, 14.20.Jn}

\maketitle

%%%%%%%%%%%%%%%%%%%%%%%%%%%%%%%%%%%%%%%%%%%%%%%%%%%%%%%%%%%%%%%%%%%%%%%%%
% INTRODUCTION
%%%%%%%%%%%%%%%%%%%%%%%%%%%%%%%%%%%%%%%%%%%%%%%%%%%%%%%%%%%%%%%%%%%%%%%%%
%\section{Introduction}

One of the open questions in fundamental particle physics is to look
for baryon number violation in nature~\cite{proton1,proton2}, which
is key point to understand the observed matter anti-matter
asymmetry. There are a few reasons to believe that baryon number
symmetry may not be exact symmetry. This is because of the three
conditions for generating this asymmetry pointed out originally by
Sakharov in 1967~\cite{sakharov}: (a) existence of $CP$-violation;
(b)baryon number violating interactions, and (c) the presence of out
of thermal equilibrium conditions in early universe. If indeed such
interactions are there, the important question is how one can
observe them in experiments. In 1980, it was pointed out by Marshak
\cite{marshak} that a crucial test of baryon number violation is
neutron-antineutron ($N-\overline{N}$) oscillation. After this
proposal was made, many experiments had been carried out for
searching $N-\overline{N}$ oscillation~\cite{nn1}. The last
experiment in the free neutron system at the ILL sets an upper limit
of $8.6\times 10^7$ s (90\% confidence level) on the oscillation
time~\cite{ill}.\\
%Two interesting processes of experimental interest are (i) proton
%decay, e.g., $p \rightarrow e^+ +\pi^0$, $\overline{\nu} + K^0$,
%etc.~\cite{proton1,proton2}, and (ii) $N -\overline{N}$
%oscillation~\cite{nn1,nn2}.

\begin{figure}[htb]
\centering
\includegraphics[width=7cm]{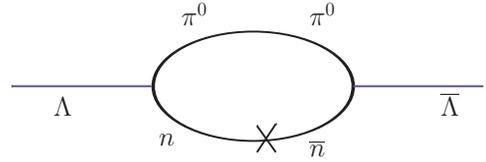}
\caption[9pt]{Demonstration of
$\Lambda-\overline{\Lambda}$\,oscillation induced by
$n-\bar{n}$\,oscillation as proposed in Ref.~\cite{luk}.}
\label{fig:plot}
\end{figure}

 It is worth noting that if $N -\overline{N}$ oscillation exists,
then $\Lam -\Lamb$ oscillation can take place as well as indicated
in Fig.~\ref{fig:plot}, which was firstly proposed by
K.-B.~Luk~\cite{luk}.  In this paper, we would like to consider the
phenomenology of $\Lam- \Lamb$ oscillation for both free $\Lam$ and
effect of an external field on $\Lam$ baryon, in particular, the
effect of an external magnetic field on the opposite magnetic
moments of $\Lam$ and $\Lamb$. Moreover, we firstly propose to
search for the $\Lam-\Lamb$ oscillation in the coherent production
in $J/\psi \rightarrow \Lam\Lamb$ decay. We discuss the observable
for both time-dependent and time-independent correlated production
rate.

%\section{Framework for $\Lam-\Lamb$ Oscillations}

The time evolution of the $\Lam -\Lamb$ oscillation is described by
the Schr\"odinger-like equation as
\begin{equation}
i \frac{\partial}{\partial t}\,\left(
\begin{array}{c}
\Lam(t)\\
\Lamb(t)
\end{array}\right)
={\bf M} \,\left(
\begin{array}{c}
\Lam(t)\\
\Lamb(t)
\end{array}\right),
\label{eq:Schrodinger}
\end{equation}
where the ${\bf M}$ matrix is Hermitian and defined as
\begin{equation}
{\bf M} = \left(
\begin{array}{cc}
m_{\Lam} - \Delta E_{\Lam} & \delta m_{\Lam \Lamb}\\
\delta m_{\Lam \Lamb} & m_{\Lamb} - \Delta E_{\Lamb}\\
\end{array}\right).
\label{eq:operator}
\end{equation}
Here $\delta m$ is the $\Delta B =2$ transition mass between $\Lam$
and $\Lamb$, $m_{\Lam}$ ($m_{\Lamb}$) is the mass of $\Lam$
($\Lamb$) baryon, and $\Delta E_{\Lam}= - \vec{\mu}_{\Lam} \cdot
\vec{B}$ and $\Delta E_{\Lamb}= - \vec{\mu}_{\Lamb} \cdot \vec{B}$
 are energy split due to external field $\vec{B}$.
 $\vec{\mu}_{\Lam, \Lamb}$ is magnetic moment of $\Lam$, $\mu_{\Lam} = -\mu_{\Lamb} = -
 0.613\mu_N$.
($\mu_N=3.152 \times 10^{-14} MeV\cdot T^{-1}$ is nuclear magneton.)

For produced unbound $\Lam$ propagating in vacuum without external
field, both $\Delta E_{\Lam}$ and $\Delta E_{\Lamb}$ equal to zero.
$CPT$ invariance imposes $m_{\Lam} \equiv m_{\Lamb}$ and $\Delta
E_{\Lam} = - \Delta E_{\Lamb}$. The equality of the off-diagonal
elements follows from $CP$ invariance. The two eigenstates $|\Lam_H
\rangle$ and $|\Lam_L\rangle$ of the effective Hamiltonian matrix
${\bf M}$ are given by
\begin{eqnarray}
|\Lam_H \rangle = \frac{1}{\sqrt{2}}(\sqrt{1+z} |\Lam \rangle +
\sqrt{1-z} |\Lamb \rangle ) ,  \nonumber \\
|\Lam_L \rangle = \frac{1}{\sqrt{2}}(\sqrt{1-z} |\Lam \rangle -
\sqrt{1+z}  |\Lamb \rangle )\, , \label{eq:eigenstates}
\end{eqnarray}
where $z = \frac{2 \Delta E}{\Delta m}$,  $\Delta E = |\Delta
E_{\Lam}| = |\Delta E_{\Lamb}|$ and $\Delta m \equiv m_H - m_L = 2
\sqrt{(\Delta E)^2 + \delta m^2_{\Lam \Lamb}}$ with $m_H$ ($m_L$)
denoting the mass of ``heavy (H)"  $\Lam_H$ (``light (L)" $\Lam_L$)
baryon. In the absence of external field, one has $\Delta E = 0$,
thus one gets $z =0$. While assuming that $\delta m_{\Lam\Lamb} \sim
\delta m_{n\overline{n}}$ for the first order, we have $\delta
m_{\Lam\Lamb} < 10^{-23} $ eV. It indicates that an external field
will make $\Delta m \gg \delta m$,  as a result we have $z
\rightarrow 1$.

From Eq.~\eqref{eq:eigenstates}, the corresponding eigenvalues are
\begin{equation}
\lambda_{\Lam_H} = m_{\Lam} + \sqrt{(\Delta E)^2 + \delta
m^2_{\Lam\Lamb}}\,,
 \label{eq:eigenvalue1}
\end{equation}
\begin{equation}
\lambda_{\Lam_L} = m_{\Lam} - \sqrt{(\Delta E)^2 + \delta
 m^2_{\Lam \Lamb} }\, . \label{eq:eigenvalue2}
\end{equation}
 Thus, starting with a beam of $\Lam$, the probability of generating a
$\Lamb$ after time $t$, ${\cal P}(\Lamb, t)$, is described by the
following equation
\begin{equation}
 {\cal P}(\Lamb, t) = \frac{\delta m^2_{\Lam \Lamb}}{\delta m_{\Lam \Lamb}^2
  + (\Delta E)^2}\mbox {sin}^2(\sqrt{\delta m^2_{\Lam \Lamb} + (\Delta
  E)^2} \cdot t). \label{eq:prob}
\end{equation}
%%% add some discussion on free and quasi-free
For free $\Lam$, we have $\Delta E =0$, and Eq.~\eqref{eq:prob}
becomes
\begin{equation}
 {\cal P}(\Lamb, t) = \mbox{sin}^2(\delta m_{\Lam \Lamb} \cdot t). \label{eq:probfree}
\end{equation}

Hereafter, we consider the possible search of $\Lam -\Lamb$
oscillation  in the $J/\psi \rightarrow \Lam\Lamb$ decay, in which
the coherent $\Lam \Lamb$ events are generated with strong boost.
Here we assume that possible strong multiquark effects that involve
seaquarks play no role in $J/\psi \to \Lam \Lamb$
decays~\cite{voloshin}.  The amplitude for $J/\psi$ decaying to
$\Lam \Lamb$ is $\langle \Lam \Lamb |{\cal H}| J/\psi \rangle$, and
the $\Lam \Lamb$ pair system is in a state with charge parity $C=
-1$, which can be defined as
\begin{equation}
|\Lam \Lamb \rangle^{C=-1} = \chi_1 \frac{1}{\sqrt{2}} \left [ |\Lam
\rangle |\Lamb\rangle -|\Lamb\rangle |\Lam\rangle\right ]\,,
\label{eq:lamantilam}
\end{equation}
where $\chi_1$ is the symmetric spin triplet and will be suppressed
since it does not affect our calculations \cite{Kang}.

We shall analyze the time-evolution of $\Lam \Lamb$ system produced
in $J/\psi$ decay. Following the $J/\psi \rightarrow \Lam\Lamb$
decay, the $\Lam$ and $\Lamb$ will go separately and the proper-time
evolution of the particle states
$|\Lam_{\small\mbox{phys}}(t)\rangle$ and
$|\Lamb_{\small\mbox{phys}}(t)\rangle$ are given by
\begin{eqnarray}
|\Lam_{\small\mbox{phys}}(t) \rangle & = & (g_+(t) + z g_-(t)
|\Lam \rangle + \sqrt{1-z^2} g_-(t) |\Lamb \rangle, \nonumber \\
|\Lamb_{\small\mbox{phys}}(t) \rangle & = & (z g_-(t) -g_+(t))
|\Lamb \rangle - \sqrt{1-z^2} g_-(t) |\Lam \rangle, \nonumber \\
\label{eq:lam_time}
\end{eqnarray}
where
\begin{eqnarray}
g_{\pm} = \frac{1}{2} (e^{-im_H t-\frac{1}{2} \Gamma_H t} \pm
e^{-im_L t - \frac{1}{2}\Gamma_L t} ),
 \label{eq:define}
\end{eqnarray}
%with $m = \frac{m_H + m_L}{2}$.
with definitions
\begin{eqnarray}
m &\equiv& \frac{m_L + m_H}{2}, \, \, \Delta m \equiv m_H - m_L,
\nonumber \\
\Gamma &\equiv & \frac{\Gamma_L + \Gamma_H}{2}, \, \Delta \Gamma
\equiv \Gamma_H - \Gamma_L,
 \label{eq:define2}
\end{eqnarray}
Note that here, $\Delta m$ is positive by definition, while the sign
of $\Delta \Gamma$ is to be determined by experiments.

In practice, one define the following oscillation parameters in a
similar fashion as in neutral $B$ and $D$ mixing cases:
\begin{eqnarray}
 x_{\Lam} \equiv \frac{\Delta m}{\Gamma}, \, y_{\Lam} \equiv \frac{\Delta
 \Gamma}{2\Gamma}.
\label{eq:define3}
\end{eqnarray}
Then we consider a $\Lam \Lamb$ pair in $J/\psi$ decay with definite
charge-conjugation eigenvalue. The time-dependent wave function of
$\Lam \Lamb$ system with $C=-1$ can be written as
\begin{eqnarray}
|\Lam\Lamb (t_1,t_2) \rangle &=& \frac{1}{\sqrt{2}} [
|\Lam_{\small\mbox{phys}}({\bf k_1},t_1)
\rangle|\Lamb_{\small\mbox{phys}}({\bf k_2}, t_2) \rangle \nonumber \\
 & - & |\Lamb_{\small\mbox{phys}}({\bf k_1}, t_1) \rangle |\Lam_{\small\mbox{phys}}({\bf k_2}, t_2) \rangle ],
 \label{eq:lamlamb_time}
\end{eqnarray}
where ${\bf k_1}$ and ${\bf k_2}$ are the three-momentum vector of
the two $\Lam$ baryons.  We now consider decays of these correlated
system into various final states. The amplitude of such joint
decays, one $\Lam$ decaying to a final state $f_1$ at proper time
$t_1$, and the other $\Lam$ to $f_2$ at proper time $t_2$, is given
by
\begin{eqnarray}
A(J/\psi &\to& \Lam_{\small\mbox{phys}}\Lamb_{\small\mbox{phys}}
\to f_1 f_2) \equiv \frac{1}{\sqrt{2}}\times \nonumber \\
&& \{ [g_+(t_1)g_-(t_2) - g_-(t_1)g_+(t_2)] a_2 - \nonumber \\
&& [g_+(t_1)g_+(t_2)-g_-(t_1)g_-(t_2) ] a_1\},
 \label{eq:amp}
\end{eqnarray}
where
\begin{eqnarray}
a_1 &\equiv & A_{f_1}\overline{A}_{f_2} - \overline{A}_{f_1}A_{f_2}
= A_{f_1}A_{f_2} (\lambda_{f_2} - \lambda_{f_1}),
\nonumber \\
 a_2 &\equiv& z(A_{f_1} \overline{A}_{f_2} +
\overline{A}_{f_1} A_{f_2}) - \sqrt{1-z^2}(A_{f_1} A_{f_2} -
\overline{A}_{f_1}\overline{A}_{f_2}) \nonumber \\
 &=& A_{f_1}A_{f_2} [ z(\lambda_{f_2} + \lambda_{f_1}) -
 \sqrt{1-z^2}(1-\lambda_{f_1}\lambda_{f_2})],
 \label{eq:aa}
\end{eqnarray}
with $A_{f_i} \equiv \langle f_i|{\cal H}|\Lam\rangle$,
$\overline{A}_{f_i} \equiv \langle f_i|{\cal H}|\Lamb\rangle$ ($i
=1,2$), and define
\begin{eqnarray}
\lambda_{f_i} \equiv
 \frac{\langle f_i | {\cal H}|\Lamb\rangle}{\langle f_i|{\cal
 H}|\Lam\rangle} =\frac{\overline{A}_{f_i}}{A_{f_i}},
\label{eq:lambda_define}
\end{eqnarray}
\begin{eqnarray}
\overline{\lambda}_{\overline{f}_i} \equiv
 \frac{\langle \overline{f_i} | {\cal H}|\Lam\rangle}{\langle \overline{f_i}|{\cal
 H}|\Lamb\rangle}
 =\frac{A_{\overline{f_i}}}{\overline{A}_{\overline{f_i}}}.
\label{eq:lambda_define2}
\end{eqnarray}

 In the process $e^+e^- \to J/\psi \to \Lam\Lamb$, the $\Lam\Lamb$ pairs are strongly boosted,
 so that the decay-time difference (t=$\Delta
t_- = (t_2 - t_1)$) between $\Lam_{\small\mbox{phys}} \to f_1$ and
$\Lamb_{\small\mbox{phys}} \to f_2$ can be measured easily. From
Eq.~\eqref{eq:amp},  one can derive the general expression for the
time-dependent decay rate
\begin{eqnarray}
&&\frac{d\Gamma(J/\psi\to
\Lam_{\small\mbox{phys}}\Lamb_{\small\mbox{phys}} \to f_1
f_2)}{dt} =  {\cal N} e^{-\Gamma|t|}\times\nonumber \\
&&[(|a_1|^2+|a_2|^2)\mbox{cosh}(y\Gamma t) + (|a_1|^2 -
|a_2|^2)\mbox{cos}(x\Gamma t) \nonumber \\
&& +2{\cal R}e(a_1 a^*_2)\mbox{sinh}(y\Gamma t) +2{\cal I}m(a_1
a^*_2)\mbox{sin}(x\Gamma t)],
 \label{eq:decay_rate}
\end{eqnarray}
where ${\cal N}$ is a common normalization factor. In
Eq.~\eqref{eq:decay_rate}, terms proportional to $|a_1|^2$ are
associated with decays that occur without any net oscillation, while
terms proportional to $|a_2|^2$ are associated with decays following
a net oscillation. The other terms are associated with the
interference between these two cases. In the following discussion,
we define
\begin{eqnarray}
R(f_1,f_2; t) \equiv \frac{d\Gamma(J/\psi \to
\Lam_{\small\mbox{phys}}\Lamb_{\small\mbox{phys}} \to f_1 f_2)}{dt}.
 \label{eq:decay_rate_define}
\end{eqnarray}

 For a given state $f_1 f_2 = (p\pi^-)
(p\pi^-)$, we have $a_1 = 0$ and $a_2 = 2 A_{p\pi^-}
\overline{A}_{p\pi^-}$. Thus one can write $R(p\pi^-,p\pi^-; t)$ as
\begin{eqnarray}
  R(p\pi^-,p\pi^-; t) = {\cal N}
\frac{1}{4} e^{-\Gamma| t|}|a_2|^2 [\mbox{cosh}(y\Gamma t)
  -\mbox{cos}(x\Gamma t)]. \nonumber \\
 \label{eq:decay_ratesim}
\end{eqnarray}

At BES-III experiment, the external magnetic field is about 1.0
Tesla, in which case, $\Delta E \sim 2\times 10^{-11}$ MeV, thus $z
\sim 1$.  Taking into account that $|\lambda|$,
$|\overline{\lambda}| \ll 1$ and $x_{\Lam}$, $y_{\Lam} \ll 1$ and $z
\rightarrow 1.0$, expanding the time-dependent for $x t$, $y t$ up
to order $x_{\Lam}^2$, and $y^2_{\Lam}$ with neglecting $CP$
violation in the expressions, we can write Eq.
\eqref{eq:decay_ratesim} as
\begin{eqnarray}
  R(p\pi^-,p\pi^-; t) = {\cal N}
 e^{-\Gamma| t|} |A_{p\pi^-}|^2 |\overline{A}_{p\pi^-}|^2
 \frac{x_{\Lam}^2 +y^2_{\Lam}}{2} (\Gamma t)^2. \nonumber \\
 \label{eq:decay_rate1}
\end{eqnarray}

For $f_1 f_2 =(p\pi^-)(\overline{p}\pi^+)$, we have $a_1 =
A_{p\pi^-}\overline{A}_{\overline{p}\pi^+}
(1-\lambda_{p\pi^-}\overline{\lambda}_{\overline{p}\pi^+})$ and $a_2
= A_{p\pi^-}\overline{A}_{\overline{p}\pi^+}
(1+\lambda_{p\pi^-}\overline{\lambda}_{\overline{p}\pi^+})$. Thus
time-dependent decay rate can be expressed as
\begin{eqnarray}
  R(p\pi^-,\overline{p}\pi^+; t) &=& {\cal N} \frac{1}{2}
 e^{-\Gamma| t|} |A_{p\pi^-}|^2 |\overline{A}_{p\pi^-}|^2
 (1+ y \Gamma t) \nonumber \\
 &\approx &{\cal N} \frac{1}{2}
 e^{-\Gamma| t|} |A_{p\pi^-}|^2 |\overline{A}_{p\pi^-}|^2.
 \label{eq:decay_rate2}
\end{eqnarray}

We define the following observable
\begin{eqnarray}
{\cal R}(t) \equiv  \frac{ R(p\pi^-,p\pi^-; t) +
R(\overline{p}\pi^+,\overline{p}\pi^+;
t)}{R(p\pi^-,\overline{p}\pi^+; t)+ R(\overline{p}\pi^+,p\pi^-; t)}
 \label{eq:cptime}
\end{eqnarray}
Combining Eqs.~\eqref{eq:decay_rate1} and~\eqref{eq:decay_rate2},
 one obtains
\begin{eqnarray}
{\cal R}(t) =  2|\lambda_{p\pi^-}|^2 \frac{x^2_{\Lam}
+y^2_{\Lam}}{2} (\Gamma t)^2.
 \label{eq:cptime-z}
\end{eqnarray}

For completeness, we derive general expressions for time-integrated
decay rates into a pair of final states $f_1$ and $f_2$
\begin{eqnarray}
 R(f_1,f_2)  &= &\frac{1}{4}{\cal N}
 \big[(|a_1|^2+|a_2|^2)\frac{1}{1-y^2}\nonumber \\
    && +(|a_1|^2-|a_2|^2)\frac{1}{1+x^2}\big].
 \label{eq:cp_independ}
\end{eqnarray}
At last,the ratio of two probabilities mentioned above can be
rewritten as
\begin{equation}
{\cal R} \equiv \frac{ R(p\pi^-,p\pi^-) +
R(\overline{p}\pi^+,\overline{p}\pi^+)}{R(p\pi^-,\overline{p}\pi^+)+
R(\overline{p}\pi^+,p\pi^-)} = 2 |\lambda_{p\pi^-}|^2
(x_{\Lam}^2+y_{\Lam}^2) . \label{eq:inter}
\end{equation}

If there is no external field and the $\Lam$ is free, we have $z=0$,
Eq.~\eqref{eq:cptime-z} becomes
\begin{eqnarray}
{\cal R}(t) =  \frac{1}{2} \frac{x^2_{\Lam} +y^2_{\Lam}}{2} (\Gamma
t)^2,
 \label{eq:cptimez0}
\end{eqnarray}
and the time-independent ratio in Eq.~\eqref{eq:inter} becomes
\begin{eqnarray}
{\cal R} =  \frac{x^2_{\Lam} +y^2_{\Lam}}{2}.
 \label{eq:cpintez0}
\end{eqnarray}
We assume that $y = 0$, thus Eq.~\eqref{eq:cpintez0} can be written
as ${\cal R } \approx  x^2_{\Lam}/2 = 2\delta m^2/\Gamma^2 $.

 In experiment at BES-III, about $10\times 10^9$ $J/\psi$ and
$3\times 10^{9}$ $\psi(2S)$ data samples can be collected per year's
running according to the designed luminosity of BEPCII in
Beijing~\cite{besiii,bepcii}. Assuming that no signal events of
$J/\psi \rightarrow \Lam_H \Lam_L \rightarrow ( p\pi^-)( p\pi^-)$ or
$(\overline{p}\pi^+)(\overline{p}\pi^+)$ are observed, we can set an
upper limit of $\delta m < 10^{-15}$ MeV at 90\% confidence level.
It will be the first search of $\Lam -\Lamb$ oscillation
experimentally. In the future, at the next generation of
$\tau$-charm factory with $10^{35}$ luminosity, the expected
sensitivity of measurement of $\Lam-\Lamb$ oscillation would be
$\delta m < 10^{-17}$ MeV at 90\% confidence level.

It is known that one has to fit the proper-time distribution as
described in Eq.~\eqref{eq:cptime-z} in experiments to extract the
$\Lam$ oscillation parameters.  At a symmetric $\psi$ factory,
namely, the $J/\psi$ is at rest in the Central-Mass (CM) frame.
Then, the proper-time interval between the two $\Lam$ baryons is
calculated as
\begin{eqnarray}
  \Delta t = (r_{\Lam} -r_{\Lamb})\frac{m_{\Lam}}{c{\bf |P|}},
 \label{eq:propertime}
\end{eqnarray}
where $r_{\Lam}$ and $r_{\Lamb}$ are $\Lam$ and $\Lamb$ decay
length, respectively, and ${\bf P}$ is the three-momentum vector of
$\Lam$. Since the momentum can be calculated with $J/\psi$ decay in
the CM frame, all the joint $\Lam\Lamb$ decays in this paper can be
used to study $\Lam-\Lamb$ oscillation  in the symmetric $J/\psi$
factory.

The average decay length of the $\Lam$ baryon in the rest frame of
$J/\psi $ is $c\tau_{\Lam} \times (\beta \gamma)_{\Lam} \approx 7.6$
cm. At BES-III,  the impact parameter resolution of main draft
chamber, which are directly related to decay vertex resolution of
$\Lam$, are described in Ref.~\cite{besiii}, from which we can get
that the resolution for the reconstructed $\Lam$ decay length should
be less than $200 \mu$m within the coverage of the detector. This
means BES-III detector is good enough to separate the two $\Lam$
decay vertices, so that the oscillation parameters can be measured
by using time information.

In conclusion, if $N-\overline{N}$ oscillation exists, then it would
be possible to induce $\Lam-\Lamb$ oscillation. We suggest that the
coherent $\Lam\Lamb$ events from the decay of $J/\psi \rightarrow
\Lam\Lamb$ can be used to search for possible $\Lam -\Lamb$
oscillation. The $\Lam$ baryons from the $J/\psi$ decay are strongly
boosted, so that it will offer the possibility to measure the
proper-time interval $\Delta t$ between the fully reconstructed
$\Lam$ and $\Lamb$. Both coherent and time-dependent information can
be used to extract $\Lam-\Lamb$ oscillation parameter. With one
year's luminosity at BES-III, we can set an upper limit of $\delta m
< 10^{-15}$ MeV at 90\% confidence level, corresponding to about
$10^{-6}$ s of $\Lam-\Lamb$ oscillation time. It will be the first
search of $\Lam -\Lamb$ oscillation experimentally.  It is important
to note that both $\Lam$ baryon decays can be fully reconstructed
without dilution from background since second decay vertex of
$\Lam$, particle identification will strongly suppress backgrounds.
The BES-III experiment is collecting data at $J/\psi$ peak now, and
we expect to see the first result of $\Lam-\Lamb$ oscillation
sooner. The next generation of $\tau$-charm factory will be
important to search for this kind of new physics.  Finally, we have
to address that the shorter mean life of $\Lam$ can significantly
hamper a sensitive search for the $\Lam-\Lamb$ oscillation as stated
in Ref~\cite{luk}.
%\section*{Acknowledgments}

%%%%%%%%%%%%%%%%%%%%%%%%%%%%%%%%%%%%%%%%%%%%%%%%%%%%%%%%%%%%%%%%%%%%%%%%%
% BIBLIOGRAPHY
%%%%%%%%%%%%%%%%%%%%%%%%%%%%%%%%%%%%%%%%%%%%%%%%%%%%%%%%%%%%%%%%%%%%%%%%%

%\bibliographystyle{h-physrev2-original}   %

\end{document}